\documentclass[lettersize,journal]{IEEEtran}
\usepackage{amsmath,amsfonts}
\usepackage{algorithmic}
\usepackage{algorithm}
\usepackage{array}
\usepackage[caption=false,font=normalsize,labelfont=sf,textfont=sf]{subfig}
\usepackage{textcomp}
\usepackage{stfloats}
\usepackage{url}
\usepackage{verbatim}
\usepackage{graphicx}
\usepackage{booktabs}

\usepackage{xcolor}		
\usepackage[colorlinks = true,
            linkcolor = purple,
            urlcolor  = blue,
            citecolor = cyan,
            anchorcolor = black]{hyperref}	

\usepackage{graphicx}

\usepackage{multirow}
\usepackage{cite}

\usepackage{color,soul}

\begin{document}

\title{A Contention-Free Model for Converged Kubernetes on HPC}

\author{Vanessa Sochat, David Fox, and Daniel Milroy \\~\IEEEmembership{Lawrence Livermore National Laboratory}
\thanks{Livermore, CA}
\thanks{}}

\markboth{Journal of \LaTeX\ Class Files,~Vol.~14, No.~8, August~2021}%
{Shell \MakeLowercase{\textit{et al.}}: A Sample Article Using IEEEtran.cls for IEEE Journals}

\newcommand{\squeezeup}{\vspace{-5mm}}


\maketitle

\begin{abstract}
Next-generation heterogeneous, complex scientific workloads that require integration of service orchestration, simulation, and automated management of state are not well supported by traditional high performance computing (HPC). While alternative cloud options offer automation, portability, and orchestration, cloud does not generally deliver the network latency performance, fine-grained resource mapping, or application scalability that tightly-coupled HPC simulations can require. These novel requirements call for change in workflow software or design and in the underlying supporting infrastructure. Converged computing is an emerging paradigm that represents the best of both worlds of cloud and high performance computing (HPC) technology and culture, and such a paradigm offers a solution to this predicament. In this paper, we introduce a new model for compute resource co-management—an HPC workload manager, Flux Framework, running seamlessly with user-space Kubernetes ``Usernetes" to instantiate an automated, service-supporting, modular, and portable converged architecture to on-premises HPC clusters. We present the first setup of this kind, and experiments that evaluate HPC application and network performance, finding equivalent performance for bandwidth, and an MPI application that heavily utilizes CPU. We provide a reproducible experimental and deployment setup to enable the larger community to utilize and further collaborate on the work.
\end{abstract}


\section{Introduction}

The landscape of high performance computing (HPC) is changing. Applications and workflows are becoming more complex and heterogeneous to reflect the expanding set of techniques that are used to conduct science. Early parallel computing in HPC was exemplified by embarrassingly parallel \cite{Wikipedia_contributors2024-ch}, bulk synchronous parallel~\cite{Wikipedia_contributors2024-bsp}, or distributed \cite{Wikipedia_contributors2023-tv,Culler1993-zd} models. Reflecting the increasing interdisciplinary nature of computational science, modern scientific workflows extend beyond early designs to integrate pipelined simulations, ensemble methods, coupled multiscale models, in-situ visualization, AI/ML training or surrogate model inference, data movement and staging, etc. The components can be steps represented in a directed acyclic graph (DAG) and sent to a workload manager with a queue and scheduler for resource allocation and task management. Representing the steps in a DAG enables encoding complex task relationships into a well-defined data structure.
This task falls under the jurisdiction of a workflow tool, and increasingly the units of work required to complete the entire workload include services such as databases and task queues, steps with multiple components working together, and time scales that can vary from millisecond execution to AI/ML training that can take weeks or more. 

These complex workflows also require elasticity or resource dynamism, which is the ability for the workflow tool to make a decision to dynamically scale up or down based on a result or workflow need. This is especially important for cloud environments where resources might be very expensive. For example, a workflow that needs GPU resources for a small period of time should not need to provision the GPUs for the entire workflow run, but rather should be able to request, provision, use, and cleanup the resources exactly when they are needed. This flexibility translates directly to cost savings to not incur additional charges for waiting or unused resources. Cloud services, by way of the scale of the companies offering them and number of developers, can quickly adjust their products and prices to match these needs. As an example, spot instances \cite{Amazon_Web_Services2022-fw} or functions as a service are lower cost, effective solutions for many small, quick tasks, and reservations that come with discounts serve the latter. A ``function as a service" (FaaS) serverless technology that enables math kernel execution on accelerators would reduce cost and potentially increase efficiency as opposed to provisioning an entire GPU virtual machine.

In the last few years machine learning (ML) and artificial intelligence (AI) has come to the forefront of interest for industry and science. While the bread and butter of HPC applications has traditionally been parallel programming models such as the message passing interface (MPI), a separate stream of tools for ML have emerged that don't always use MPI, but instead reflect the architecture on which they need to run. For example, ML training provided by software like Pytorch or Tensorflow that requires matrix operations has developed to take advantage of single large machines running GPUs \cite{Lind2019-qe}. Interestingly, the development has come full circle -- with these same models being ported back to HPC systems, but extended to use both GPU and MPI to scale across physical machines. Another design pattern that has emerged for ML, and specifically applications that require a distributed architecture, is a set of machines that adhere to a leader-worker design \cite{Jiang2021-ue} where a single machine allocates units of work that can be expanded or contracted, restarted on failure, and generally fault tolerant. This is in contrast to MPI, which is not fully fault tolerant nor elastic, except for approaches that attempt to recover failures \cite{recover-failures} or restart an entire component of the application \cite{Georgakoudis2021-wt}. The trade-off between the tightly integrated but more fragile MPI and distributed, fault tolerant and loosely integrated worker design is performance. A decision to use one technology or the other, or even choose on-premises HPC vs. cloud environment, often comes down to a choice on where to operate along this spectrum. The users of HPC systems have the desire for all these features, and it is often the case that the infrastructure and tooling simply cannot support them.

In the last few years, a community has emerged that champions converged computing -- a movement that advocates for technologies and culture that encompasses the best of both worlds between cloud and HPC \cite{Misale2022-nc,Milroy2022-pv,Guo2023-an,Park2022-oq,Sochat2024-the-flux-operator}. Convergence aspires for these traditionally separate communities to better work together, and for the technologies to follow. The convergence can solve problems on each side of the spectrum. Cloud can benefit from the performance expertise related to resource topology or optimization that has evolved from many decades of HPC research, and HPC could benefit from the modularity, portability, and orchestration that cloud offers. While several models for convergence exist that span from bursting work to external resources \cite{Gupta2013-gb} to running clusters alongside one another \cite{Souza2019-ah} to creating executors for workflows that span environments \cite{Molder2021-rq,Di_Tommaso2017-lw}, an often overlooked paradigm is one of complete immersion or embedding -- one service or tool running elegantly with the other one, in the same space, and provided as a holistic model of compute. In the context of workload managers for HPC and Kubernetes, this means using Kubernetes not as an external service or side-by-side cluster, but as an on-premises resource that runs seamlessly with the workload manager on the same physical resources. In this model, Kubernetes is running in user-space, and is available to any user on a shared, multi-tenant HPC cluster. 

In this paper, we introduce such a model -- an HPC workload manager, Flux Framework \cite{Ahn2014-ku}, running with user-space Kubernetes, or ``Usernetes." We introduce this simple, yet powerful new paradigm by way of early experiments that demonstrate the capabilities of both setups in a single environment, and provide the community with a reproducible development environment to reproduce our work and continue discussion and innovation. In Section \ref{section:method}, we start by describing the underlying components, Usernetes and Flux Framework, and many of the challenges that had to be overcome to afford our setup. We then discuss a set of experiments that can measure HPC application and network performance in this environment, and a prototype example of an application with hybrid needs that maximally utilizes the features offered by the environment. We finish with discussion about work in progress, future developments, and hopes for the community. We believe that this new paradigm holds promise to revolutionize HPC by bringing cloud technologies directly to it.
\section{Method}
\label{section:method}
\subsection{Container Orchestration}
\subsubsection{Kubernetes}

Kubernetes is a resource management and scheduling powerhouse of industry, with 71\% usage reported by Fortune 500 companies and over 90K contributors \cite{Honeypot2022-yv,noauthor_2023-do}. It has become the de-facto standard tool for container orchestration, offering a developer and application environment that is modular, configurable, and can be deployed on demand. Historically, the barrier for adoption of such an infrastructure on shared systems was the need for root daemon services, however with rootless technologies, that has changed, offering the first opportunity for the HPC community to explore the technology space.

\subsubsection{Usernetes}
\label{section:usernetes}
Usernetes grew out of the concept of an unprivileged or ``rootless" container, which was first introduced by LXC in 2014 \cite{Graber_undated-qc}. User namespaces were available since version 3.12 of the Linux kernel, and they provide a large range of user and group identifiers that can be intelligently mapped into the container, both for access permissions and resources such as network, processes, and devices. This meant that, unlike a root daemon-run container such as the original version of Docker, escaping from the container would not give the attacker root access to a system. The container runtime ``runc" followed suit, adding support for rootless in 2017 \cite{noauthor_undated-dj} and this was quickly followed by major container technologies such as Docker and Podman, among others \cite{Suda2023-qj}. It followed that rootless containers could be used in Kubernetes by way of first adding features to the kubelet \cite{noauthor_undated-so,noauthor_undated-kd} and then developing the full Usernetes project \cite{usernetes} in 2018. Generation 1 of the project, which persisted through early 2023, was a collection of bash scripts that did not require a rootless container technology, and did not support multiple hosts. One of the authors of this paper (VS) pursued this multi-host use case in early 2023, and started engagement with the project \cite{noauthor_undated-my}. Generation 2 was released in the same year \cite{noauthor_undated-fg} that harnessed fully rootless container technologies to deploy user-space Kubernetes. This development provided a new opportunity to deploy an entire Kubernetes in user-space, and alongside a traditional HPC workload manager.

\subsection{Architecture}
\subsubsection{Mitigating Resource Management Conflict}

Deploying multiple resource managers and schedulers to manage a common set of resources facilitates policy flexibility and can enable greater management scalability \cite{Verma2015-oo}. A team at Google created a taxonomy for describing relationships between multiple schedulers, identifying monolithic, two-level, and shared-state architectures as viable choices \cite{Verma2015-oo}. The authors described advantages and disadvantages of the three approaches, ultimately deciding on the shared-state architecture for their Omega scheduler. Paraphrasing their analysis, to enable multiple scheduling policies on a set of shared resources, monolithic schedulers require static resource partitioning to maintain consistency. A two-level design like Mesos ``dynamically partitions a cluster” and controls concurrency between schedulers ``pessimistically.” Mesos works well when job sizes are small relative to the size of the cluster, and when jobs are short-lived; it can deadlock when attempting to “gang” schedule MPI jobs via resource hoarding \cite{Verma2015-oo}. Shared-state designs like Omega allow each scheduler full access to all resources, optimistically controlling concurrency to mitigate resource conflicts. Conflict resolution for coupled jobs causes marked increase in conflict fraction and scheduler busyness \cite{Verma2015-oo}.

The Flux Framework provides an alternative approach: hierarchical scheduling and resource management \cite{Ahn2014-ku}. Flux is based on the principles of hierarchical bounding, meaning that a parent grants resources to its children, and instance effectiveness, signifying that each instance can be independent and is solely responsible for its resource set \cite{flux-framework-learning-guide-adoption}. A hierarchy of schedulers and resource managers does not suffer from the same throughput limitations as a monolithic scheduler: a hierarchy of Flux instances can schedule over 100 million jobs in a single workflow \cite{Ahn2020-ci} and achieve nearly 800 jobs per second scheduling throughput \cite{flux-learning-guide}.

Deploying a second resource manager and scheduler within a Flux hierarchy with instance effectiveness bounds the scope of scheduling and resource management conflict to the instance level of deployment. Thus a second resource manager and scheduler like Usernetes can run in an instance encapsulated as a job in a hierarchy, providing lifetime control, user compartmentalization, fairshare, user-scoped utilization data, and fine-grained resource management mapping that are absent or poorly supported in Usernetes and Kubernetes. 


\subsubsection{System Requirements}

Generation 2 of Usernetes uses the base image provided by Kubernetes in Docker ``Kind" \cite{kind} project, and uses kubeadm for cluster configuration \cite{kubeadm}, containerd for the container runtime interface (CRI) \cite{containerd}, runc for the OCI-standardized container spawning \cite{runc}, and Flannel as the network fabric for the container networking interface (CNI) \cite{flannel}. Supported operating systems at the time of this work included Ubuntu 22.04 (recommended), along with Rocky Linux and Alma Linux 9. Usernetes requires a rootless container technology with Docker as first preference, but Podman and nerdctl are also supported. To support rootless mode, cgroup version 2 delegation must be enabled \cite{cgroupv2}, allowing for consumption of CPU, memory, I/O and the process namespace. The br\_netfilter \cite{brnetfilter} and vxlan \cite{vxlan} network modules must be enabled to allow for inter-node communication, and network packet checking set to a less stringent mode \cite{Roudier_undated-wu}. 

One known weakness in the setup is the use of slirp4netns for network traffic, which receives packets from a container via a TAP device and extracts payloads to send via a socket \cite{Matsumoto2024-zn}. In practice, this intermediate relay of packets leads to a slowdown in network transmission, and the authors of this paper observed greater than a $2\times$ slowdown using traditional ethernet in early experimentation \cite{fosdem-bare-metal-bros}. These system-level, often security relevant changes introduce a barrier to quick prototyping on an existing bare-metal system, as it either is the case that changes need to be thoroughly vetted, or it simply is not possible for a developer with user-level permissions. A more reasonable development approach was using a testing environment with cloud virtual machines.

\subsubsection{Deployment}
\label{section:deployment}

The choice of where to deploy one or more virtual machines that are oriented to run Usernetes also presents considerable challenges, as each potential environment can have differences with respect to nuances of networking, kernel features available, or namespaces. The first successful endeavour was prototyping a setup of an HPC workload manager running alongside Usernetes in Linux Virtual Machines ``LIMA" \cite{lima-usernetes}. This was followed by a small, private on-premises cluster deployed with Ovirt \cite{Proffitt_undated-yv} that provided the first performance testing that was presented at FOSDEM in 2024 \cite{fosdem-bare-metal-bros}. 

Amazon Web Services (AWS) Elastic Compute (EC2) was used to deploy a set of 33 virtual machines, a family of instances ``hpc7g" (hpc7g.4xlarge) \cite{Amazon_Inc2024-hs} that are powered by AWS Graviton (Arm) processes, oriented toward HPC workloads in that they offer, by default, single threading, exclusive access to the full virtual machine, and a reasonable network (128 GiB memory, and 200 Gbps EFA Network bandwidth). 
A cluster placement group was used to ensure selection of nodes that are physically close together \cite{placement-group}. The Elastic Fabric Adapter (EFA) network interface \cite{Amazon_Web_Services2022-iv} was essential for the basic functioning of MPI-dependent HPC applications in this setup on EC2. Such an interface that bypasses the operating system and goes directly to a driver provides the minimal application performance needed for communication protocols such as MPI, and bypasses some of the problems of using slirp4netns. Without such an interface, many of these workloads simply don't run, or run so unreasonably slow it would be impossible to achieve any kind of result in a reasonable amount of time.

There were several key features or workarounds specific to AWS that ultimately enabled the final setup to work. 
A startup script using the AWS Python client \cite{Amazon_Web_Services2022-jf} was used to query the set of instance hostnames via a selector label to populate the Flux configuration file. 
Secondly, default health checks needed to be disabled via an arbitrary large check duration as an erroneous failed check could bring down the final cluster. 
Thirdly, the AWS load balancer requiring subnets in two or more availability zones presented a problem to Usernetes that required the addresses but could not interact across zones. The solution was to deploy the setup with two or more subnets and availability zones to pass validation, but designate the instance group to use only one. Finally, the Elastic Fabric Adapter requires the security group to have an ingress and egress rule to itself, which is not clearly documented, nor is it an intuitive or obvious thing to do.


The final environment used for this work provides an ability to scale, a network interface that supports basic HPC workloads, and can be programmatically reproduced by way of using Terraform configurations that have embedded solutions to all of the issues we mentioned above. This Terraform setup is publicly available, and forms the basis for the work in this paper \cite{flux-usernetes}. 

\subsubsection{Software}

We use the Flux Framework, a modular, hierarchical resource management and graph-based scheduling framework developed at the Lawrence Livermore National Laboratory. Flux supports scheduling dynamic resources via its graph-based representation~\cite{Patki2023-cg}, and its modular architecture facilitates integration with cloud~\cite{Milroy2022-pv,Sochat2024-the-flux-operator}. Its hierarchical resource management enables launching nested instances, which allows for high throughput and submission of over 100 million jobs \cite{Ahn2014-ku,Ahn2020-ci,flux-learning-guide}. Flux is scalable and designed to manage and schedule very large systems, such as the upcoming exascale El Capitan system \cite{noauthor_undated-et}. 

It provides several key features that make it well-suited for combination with Usernetes. In particular, having a shared filesystem is an extra setup step and incurred cost in a cloud environment, and since the workloads (Section \ref{section:experiments}) in this study did not require one, we decided to omit it. However, this choice made tasks such as sharing a file-based key from the control plane to other workers challenging. By using ``flux archive'' \cite{flux-archive} to save the key to a memory map and ``flux exec'' \cite{flux-exec} to send it to workers, the problem is solved, and a shared filesystem is not needed to share files, or even to run commands across workers. This second ability to execute commands across the cluster from a single node was required to start and connect each node in the Usernetes cluster. In early prototypes, each virtual machine needed to be connected to via a secure shell (ssh) to manually run the command. Now, with much larger clusters, the worker nodes never have to be accessed, as all interactions with them occur from the lead broker or control plane. One final bug with the Flux setup was the use of systemd. While the HPC family of AWS instances is explicitly intended for HPC use-cases and the memory limit is set to ``unlimited" systemd by default will set a ceiling, and an additional parameter ``LimitMEMLOCK=infinity'' is required in the service file for Flux to also allow for unlimited access \cite{systemd}. 

The core of the setup requires a base of virtual machines, each of which is provisioned with an HPC workload manager, Flux Framework, the system requirements for Usernetes (Section \ref{section:usernetes}), a clone of Usernetes that is ready for starting the cluster nodes, along with applications of interest. This setup affords three different models for running jobs across nodes – without Usernetes on ``bare metal,'' with Usernetes across pods and nodes, and with Usernetes, still across pods and nodes, but deploying the entirety of Flux Framework inside of the set of pods to afford the same hierarchical management. We put bare metal in quotes because a virtual machine is not truely bare metal, but the closest to it that can be achieved in cloud. For the first model, an example would be running an HPC MPI application directly with Flux Framework. For the second second model, we might use Flux to create a partition of nodes for Usernetes, and then run a multi-pod service across them such as a database or task queue. For the third model we create an equivalent partition of nodes, but then deploy the Flux Operator \cite{Sochat2024-the-flux-operator}, a Kubernetes operator that deploys the entirety of Flux Framework inside of Kubernetes, within. As it sounds, this is running an HPC application inside of Flux, inside of Usernetes, inside of Flux. This final approach would require the HPC application to be containerized, and the Flux Operator adds the entire Flux install to it on the fly. For these experiments, we took care to ensure that the Flux inside of the container matched the outer install on the virtual machines. We ensured matching versions so the applications' execution with Flux in Usernetes was as similar as possible to the same applications' execution on the virtual machine. 

For networking, although the Flux Operator uses a headless Service \cite{noauthor_undated-wi}, its performance depends on the underlying network provided by Usernetes. For the Flux Operator to take full advantage of the Elastic Fabric Adapter, a DaemonSet \cite{noauthor_undated-bu} needed to be deployed that would expose the host EFA drivers to containers that were labeled to require the ``vpc.amazonaws.com/efa'' label under resource requests and limits \cite{noauthor_undated-hz}. Although this DaemonSet is provided by AWS to deploy EFA to traditional Kubernetes clusters, by way of not being directly on their instance types, we created a custom deployment configuration that removed the selectors and would thus deploy the DaemonSet to our Usernetes nodes.

\begin{figure}[H]
  \centering
  \includegraphics[scale=0.3]{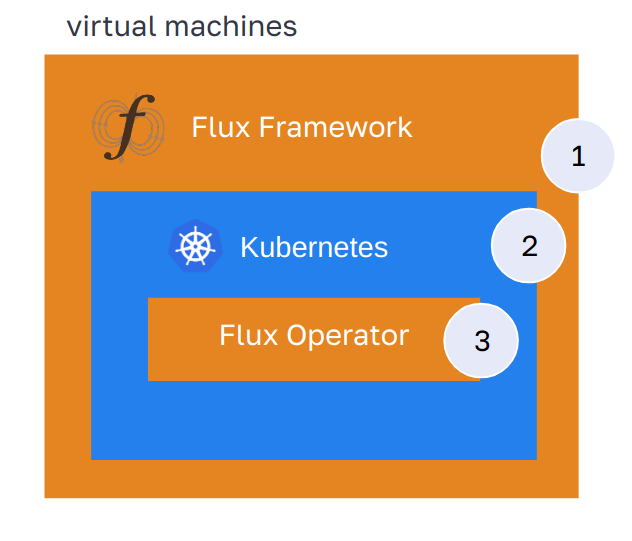}
  \caption{Flux Framework and Usernetes. The high level setup includes one or more virtual machines that are provisioned with the workload manager Flux Framework, system requirements for Usernetes, and applications of interest (1). A batch job or allocation can be created by the user to run user-space Kubernetes ``Usernetes" (2). Within a Job submit to Usernetes, by way of using the Flux Operator (3), another Flux Framework cluster is provisioned that can run the same applications.}
  \label{fig:turducken}
\end{figure}

\begin{figure}[H]
  \centering
  \includegraphics[scale=0.2]{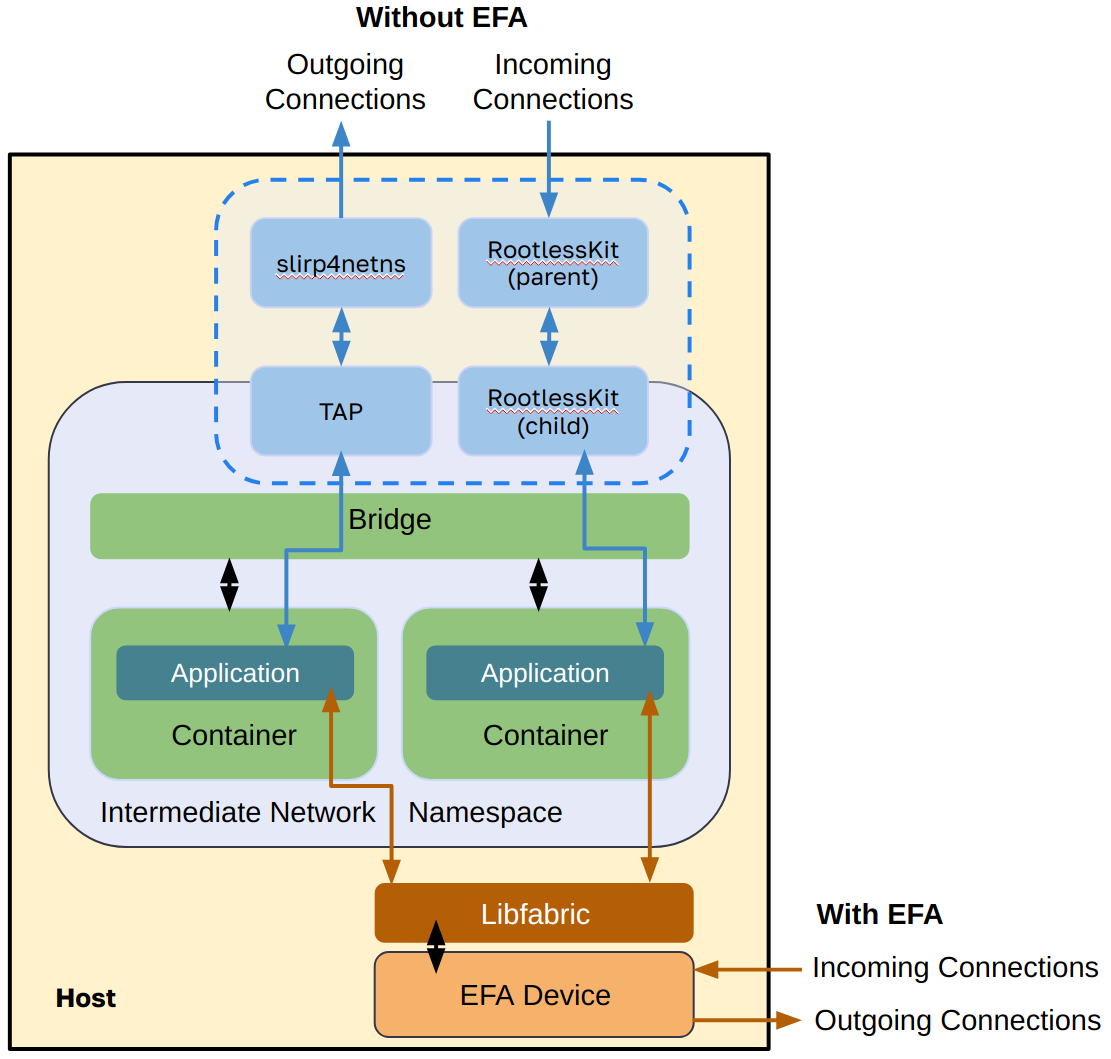}
  \caption{Networking Setup comparison between traditional user-space Kubernetes with and without the Elastic Fabric Adapter (EFA). For using Usernetes out of the box, a traditional path that uses TCP/IP is used, taking advantage of components in the blue, dotted box to afford rootless networking. This approach takes a performance hit because of the need to receive and process packets in the TAP device. The details of this approach are explained in Figure 1 of \cite{Suda2023-qj} from which the architecture is derived. By way of using EFA (bottom of diagram) applications use the user-space libfabric API to interact directly with hardware. This means that HPC applications do not take a hit as a result of the setup.}
  \label{fig:networking}
\end{figure}

A diagram of the setup with Flux Framework on the virtual machines, and Usernetes running Flux Framework within is depicted in Figure \ref{fig:turducken}, and a virtual machine level architecture is shown in Figure \ref{fig:networking}. Interestingly, this setup provides a layered architecture, where an HPC workload manager Flux is running inside of user-space Kubernetes, which is running inside of a Flux Framework allocation on the virtual machine. 

\label{section:experiments}
\subsection{Experiments}

The following experiments have two primary goals -- to assess performance issues that might arise with running traditional HPC applications in Usernetes (Section \ref{section:hpc-apps}) and to demonstrate a benefit that the setup affords -- a hybrid environment that combines an HPC workload manager and user-space Kubernetes to run HPC workloads alongside services (Section \ref{section:ml-server}).

\subsubsection{Scaling of HPC Applications}
\label{section:hpc-apps}

We are first interested in testing the strong scaling of an HPC application, namely one with MPI, across cluster sizes of 4, 8, 16, and 32. While running an MPI application in Usernetes is not strictly necessary given the availability of ``bare metal'' virtual machines, the goal of this experiment is to identify performance issues that can impact more realistic use cases such as model-parallel machine learning training or other programming models that will exercise the HPC network.

We created 33 EC2 virtual machines (Section \ref{section:deployment}) due to the need for Usernetes to have a control plane, which typically is intended to act as manager and explicitly labeled to not accept work \cite{noauthor_undated-vv}. For our HPC application experiment we first chose LAMMPS, the Large-scale Atomic/Molecular Massively Parallel Simulator \cite{LAMMPS} as a proxy application to demonstrate the ability of the setup to strong scale with MPI, and in particular for an application that requires AllReduce collective and send and receive paired calls \cite{Walker1996-bg}. A problem size of $16 \times 16 \times 8$ was chosen to account for the iterations, different scales, and final time and cost for the experiment runs. For a dedicated networking benchmark, we chose to use the OSU benchmarks \cite{Paniraja_Guptha2023-rr} to measure MPI collective performance, bandwidth, and point to point latency. We decided to test each of these application on ``bare metal'' with the HPC workload manager Flux Framework to compare with an equivalent run using the Flux Operator in Usernetes. We also chose to run a containerized variant on bare metal using Singularity \cite{Kurtzer2017-xj} to provide a point of performance comparison. We ran 20 iterations with Flux across sizes 4, 8, 16, and 32 for each of the following cases: \newline

\begin{itemize}
\item LAMMPS on bare-metal with/without a container
\item OSU Benchmarks on bare-metal with/without a container
\item LAMMPS in Usernetes using the same container
\item OSU Benchmarks in Usernetes using the same container
\item LAMMPS on bare-metal with/without container with Usernetes overhead
\item OSU Benchmarks on bare-metal with/without container, with Usernetes overhead\newline 
\end{itemize} 

In practice the above experimental setup came down to running iterations of LAMMPS and the OSU benchmarks with and without containers, then starting Usernetes, running them both in Usernetes, and then repeating the initial runs with Usernetes continuing to run to assess the potential impact of its overhead on the overall environment, and more specifically, if running Usernetes could have detrimental impact to running other applications on the system.

For experiment runs involving containers (including Usernetes) the same container was used, the difference being that ``bare metal" runs would pull it from a Docker registry to Singularity, and Usernetes would be using a rootless container that is pulled to the nodes on the first run. The same installations for each of LAMMPS and the OSU benchmarks were also built into the virtual machines. For each of the Usernetes deployments, we used the Flux Operator to replicate running the workflow using the same Flux Framework, down to the versions of dependencies and the base operating system. To ensure that all containers were loaded into the Usernetes nodes before each experiment, we ran a single, one-off run before the experiments that used the entire set of worker nodes on the cluster (N=32). This setup ensured as much similarity between environments as possible, and also allowed for a sub-experiment to assess performance of an HPC proxy application with and without a container. 

\subsubsection{Hybrid Demonstration of Environment Capabilities}
\label{section:ml-server}

\begin{figure*}
    \centering
    \includegraphics[width=\textwidth,scale=0.3]{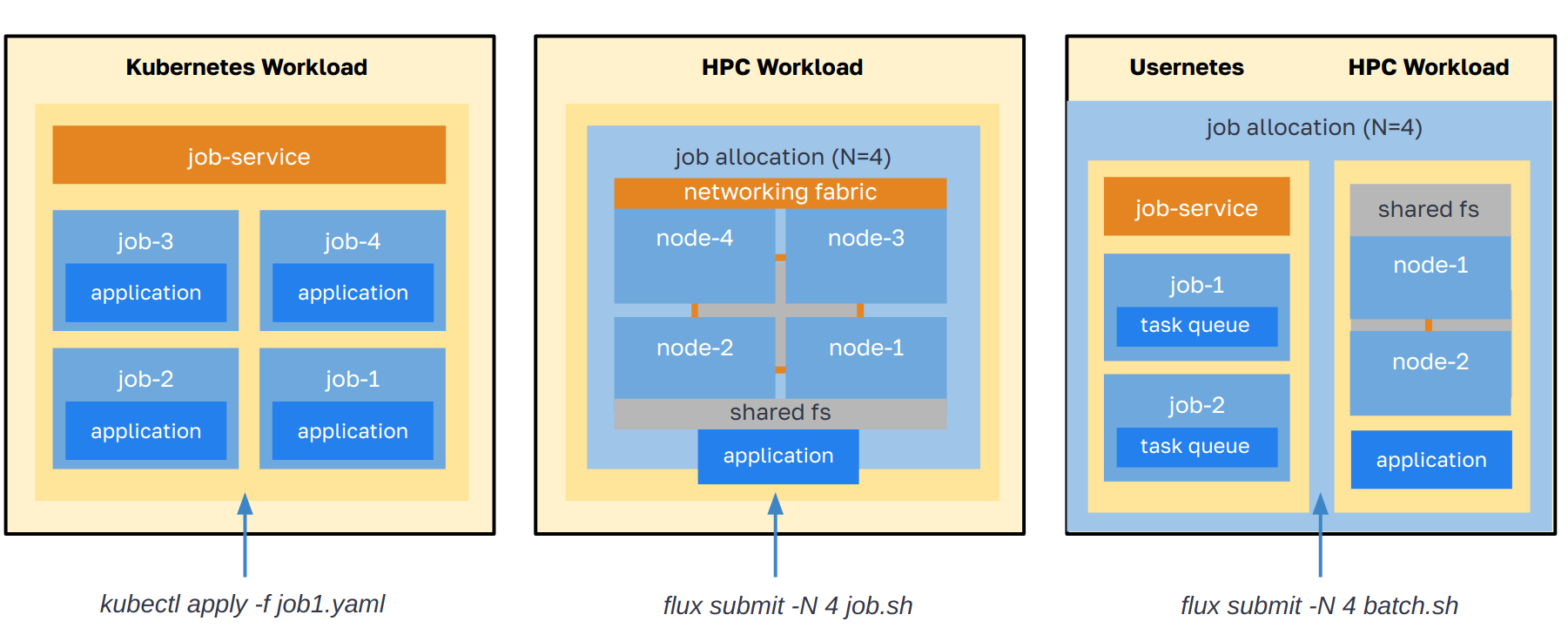}
  \caption{Running a simple, multi-node job across three setups: Kubernetes, traditional HPC, and a converged computing approach (user-space Kubernetes deployed as part of a batch job). Kubernetes (left panel) affords portability, orchestration, and modularity of services and applications at the cost of performance, reliable access to a file system, and fine-grained resource mapping. Traditional HPC workload managers (middle panel) are oriented toward applications that require tight coupling and performance, which comes at the cost of portability and easy orchestration of associated services. A converged approach with user-space Kubernetes that is deployed as part of a batch job offers the best of both worlds: bare metal HPC for performant, scaled simulation, and the modular, automated service architecture afforded by Kubernetes. Missing from this picture is the workflow tool or orchestration and scheduling of those two components.}
  \label{fig:computing-environments}
\end{figure*}

The compelling use case for a hybrid environment with an HPC workload manager and user-space Kubernetes is being able to run HPC workloads alongside services within a user-scoped allocation. To best understand this approach, we will first describe a simple setup for running a simple, multi-node application in each space (Figure \ref{fig:computing-environments}): Kubernetes, native HPC, and converged. In the Kubernetes setup, we use a ``Job" abstraction with a headless Service to give a shared network to a set of Pods, each running a container with the application inside. This approach is advantageous for the orchestrated automation and modularity of applications and services, but is designed for loosely coupled, stateful services and may not be best suited to low latency, tightly coupled applications. It would be well-suited, however, for associated services like databases, task queues, model APIs, and AI/ML components of a scientific workflow. The scheduler also is designed to support single Pods, and requires extensive modification to handle logical groups of work. Storage needs must use the container storage interface (CSI) drivers coupled with cloud solutions, and are challenging. If applications run, there is a loss of performance, and Kubernetes is not yet suited to handle fine-grained topology mapping that might be needed. 

For traditional HPC, the user submits jobs to the workload manager by way of a headless login node, and an allocation of resources is scheduled for the work. When the allocation is granted, the nodes are connected via a high speed interconnect networking fabric, and the application, which is typically built on these same machines, is run on the bare metal machines that are physically connected, and with a resource mapping that optimizes performance. Unlike Kubernetes, support for orchestration and automation is poor, with scientists often relying on workflow tools that use a shared filesystem to represent state, or bash scripts that utilize polling and custom checks. If a service such as a database or task queue is needed, it typically is a special request, as shared, open ports are not common or considered good practice for headless, private systems. However, this setup affords scalability, fine-grained resource mapping, and performance, which often are top priority for scientific simulations. 

Finally, a converged setup with a workload manager scheduling work to both bare metal HPC and user-space Kubernetes demonstrates the strength of combining the two. In this setup, user-space Kubernetes and traditional HPC simulations are running under the same allocation, and both are provisioned via a single batch job that uses a hierarchy to create sub-allocations for each. For this hypothetical example (Figure \ref{fig:computing-environments}), two of the four nodes are being used for a service that is accessible to the simulations, and the simulations run alongside the service, using it. The batch job handles the creation and cleanup of associated services along with submitting jobs for the simulations. The three setups are shown side by side in Figure \ref{fig:computing-environments}, and the advantages and disadvantages to each are shown in Table \ref{table:computing-env-features}. 

\begin{table*}[h]
  \centering
  \caption{High-level Computing Environment Features}
  \label{table:computing-env-features}
  \begin{tabular}{@{}p{2.8cm}lll@{}}
    \toprule
    & Kubernetes & HPC Workload & Converged \\
    \midrule
Application coupling & Loosely coupled & Tightly coupled & Support for both \\
Network Optimization & High Bandwidth & Low Latency &  Hardware Dependent \\
Storage & Challenging, ephemeral &  Persistent, used for state &
Persistent, not readily accessible by Usernetes\\ 
Portable  & Yes (containers, YAML) & Sometimes (with containers / workflow managers) & Yes or Sometimes \\
Scheduling Unit & Single pods or groups & Nodes, cores, down to socket or PCI bus and subsystems & Meta- or converged- schedulers \\
    \bottomrule
\end{tabular}
\end{table*}

\begin{figure*}
  \centering
  \includegraphics[scale=0.28]{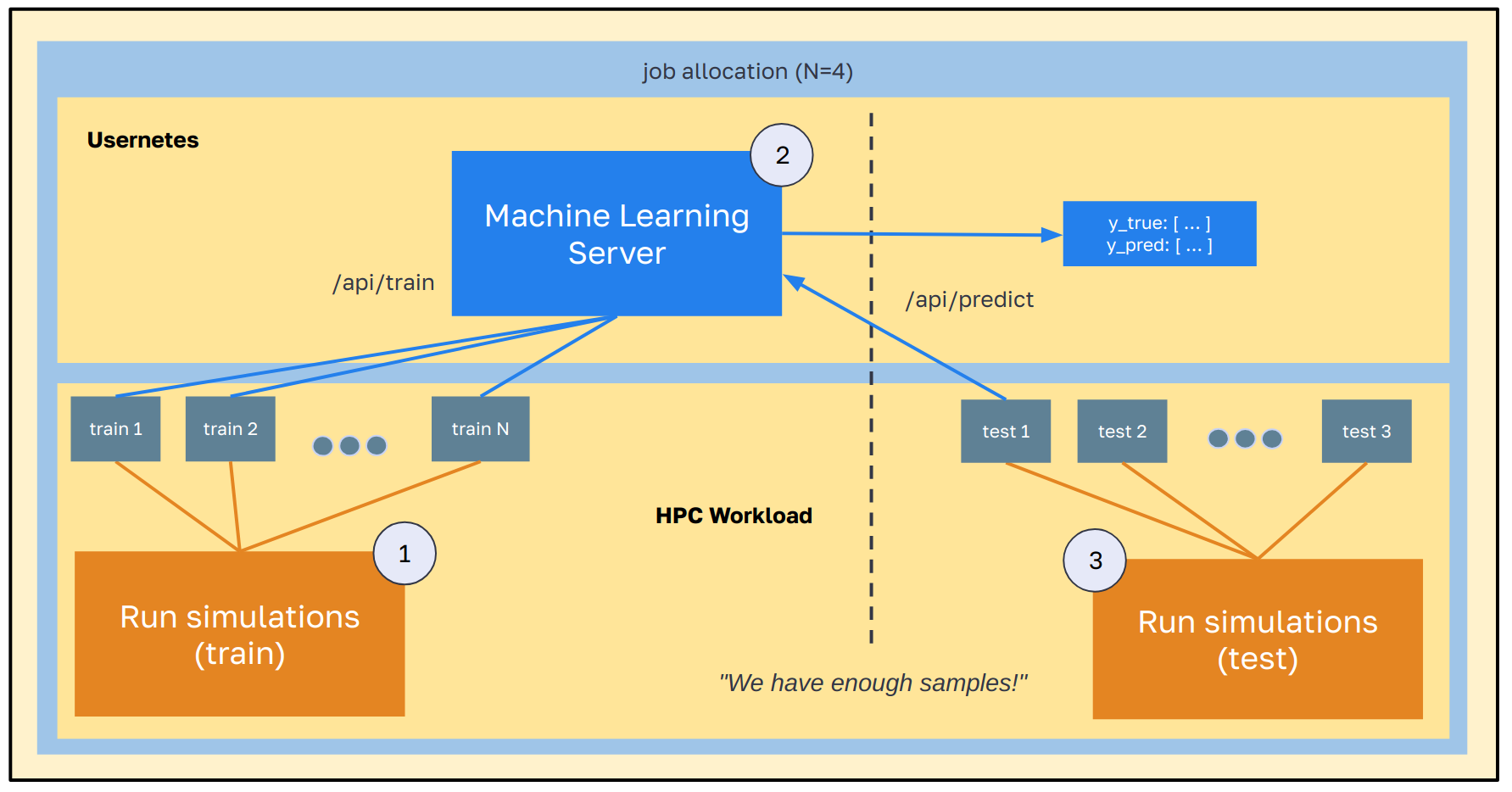}
  \caption{Hybrid machine learning and HPC workload setup. User-space Kubernetes (top panel) running alongside a traditional HPC workload manager (bottom panel) means that within the scope of a batch job, HPC simulations (in the example above, LAMMPS) can interact seamlessly with services. In this example, the entire job setup is orchestrated by a batch script that creates two sub-allocations, one for Usernetes and one for the simulations. The simulations are run by submitting jobs directly to the workload manager (1) under one sub-allocation, and each job sends results to the API exposed by the machine learning service (2). Upon completion of the initial jobs after reaching a metric of goodness, a second set of jobs are submitted (3) that only provide features and request a prediction. An assessment can then be made between predictions and actual wall times from the test runs to assess model accuracy.}
  \label{fig:ml-server}
\end{figure*}

As a multi-node job is too simple an example, we aimed to create a prototype setup with machine learning to fully demonstrate the advantages of the converged approach. Toward this goal, we developed a simple workflow that runs the same HPC application, LAMMPS, on bare metal with the workload manager Flux Framework, to predict LAMMPS wall times from the problem sizes. While the predictions themselves are not a contribution offered by this work, they are a key component of a proxy to the type of converged workload that we want to support and champion. As this setup aims to demonstrate the hybrid model, the specific problem size and number of nodes is less important than in our previous scaling study, and so we chose a cost effective cluster size of 5 nodes (allowing for LAMMPS to run on 4 nodes) and a problem size range between 1 and 8 for each of the dimensions for x, y, and z that go into LAMMPS. For this setup, LAMMPS jobs are submitted to the workload manager and on completion, the result data (the Y in the model being the LAMMPS wall times, and X features as the randomly chosen x, y, and z parameters that define a bounding box or problem size) to a machine learning service being run by Usernetes. This setup is depicted in Figure \ref{fig:ml-server}. Specifically, this figure would be a more detailed manifestation of the third panel in Figure \ref{fig:computing-environments}.

For our machine learning server, we used an approach that harnesses online machine learning \cite{river}, a paradigm that postulates samples being received by models one at a time, in a stream, as opposed to being available all at once. We used a custom server \cite{ml-server} that allows for creation, training, and testing of arbitrary models. To demonstrate the capabilities of a streaming machine learning server, we selected three regression models to train, including linear regression \cite{river-linear-regression}, bayesian linear regression \cite{bayesian-linear-regression}, and passive aggressive regression \cite{pa-regression}. While our choice of parameters and models is not a contribution of this work, we took a reasonable approach of taking a vector of features, X, to predict a value, Y. We used practices recommended by the streaming ML software, including simple preprocessing, and a standard scaler \cite{standard-scaler} to ensure the data has zero mean and unit variance. Our choice of streaming or online machine learning reproduces the behavior of real scientific workflows with dynamic ensembles, where data is generated across time in small units as opposed to being all available at once in a large matrix. For a dynamic workflow, training might continue until a metric of goodness is achieved, and this is well-suited for the streaming approach.

In practice, the hypothetical user simply needs to start the machine learning service, a Kubernetes Deployment \cite{Conventions_undated-um}, and then ensure that the port for it is exposed via Ingress \cite{noauthor_undated-me}, or directly from the port of the Pod mapped to the host. A simple script, ideally in a container for reproducibility, can then be called against the API endpoint to create the models, which in this case were three regression models. Once the models are created, another script (also ideally from a container) can submit one or more jobs directly to the workload manager that will run the HPC workload (LAMMPS) and then submit the resulting wall time (Y) and features (x, y, z) to the server training endpoint. The model trains, one point at a time, in this fashion, up until the user decides it has reached some metric of goodness or number of samples (also provided by the API server). When these submissions are done, a second batch is sent off to again run LAMMPS, but the actual wall times, Y, are left out, and the server is asked for a prediction based on the feature vector with the problem size, x, y, z. Since these jobs are actually running LAMMPS, it means that we have the true value of the wall time, and can use that to plot the actual vs. predicted times for each model. This final step can show how well the model did with an $R^2$ value. Note that although simple regression models were chosen for this demonstration, online machine learning offers many different kinds of models. For our prototype example, we did 1000 training runs and 250 test cases, each randomly selecting problem sizes for x, y, z between 1 and 8. This range was chosen to pin the maximum runtime (and thus cloud cost) to a maximum value. A more rigorous (and expensive) study would warrant analyzing runtime characters of a larger range of problem sizes. 

This demonstration, although simple, reflects a powerful hybrid setup where services provided by Usernetes run easily alongside HPC. Although this small setup did not use a workflow tool, one would be recommended for future, more complex endeavors. 

\section{Results}

\subsection{LAMMPS}

\begin{figure}
  \centering
  \includegraphics[scale=0.3]{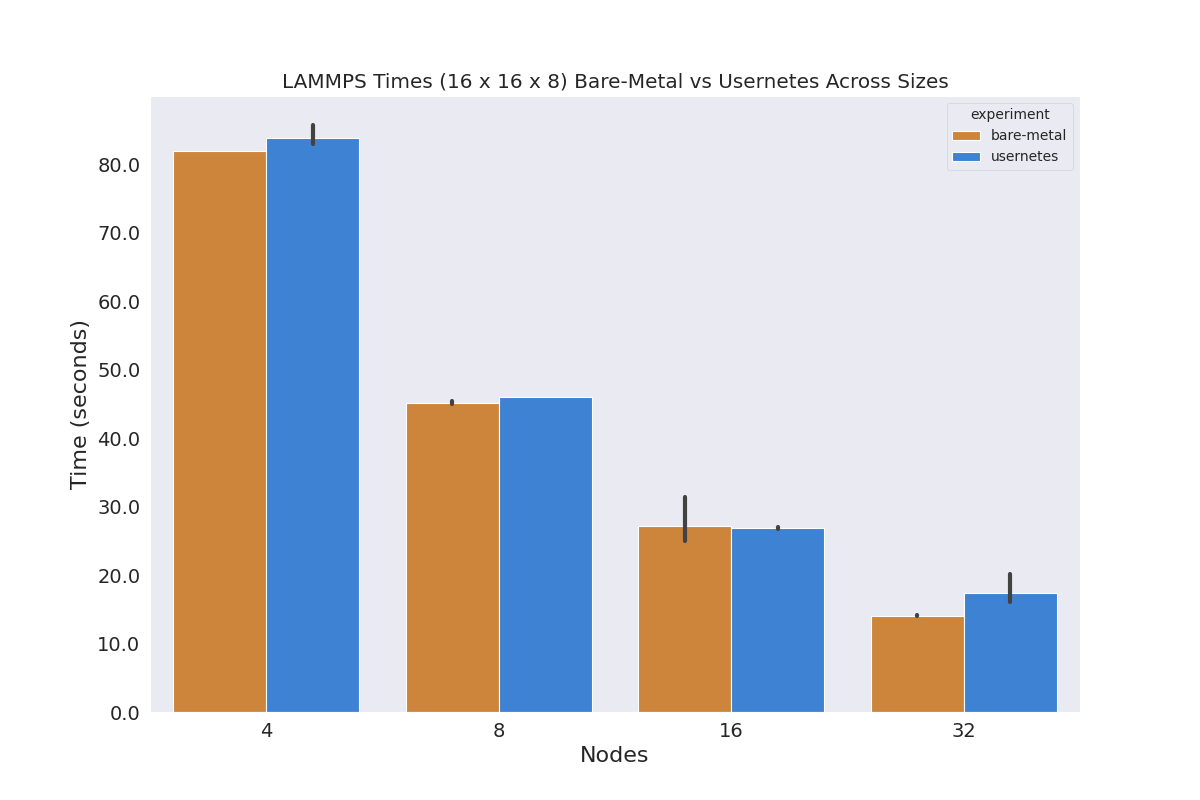}
  \caption{LAMMPS runs with strong scaling for a $16 \times 16 \times 8$ problem size and 20 iterations per experiment and size. LAMMPS running in Usernetes was only marginally slower with lower variability. Node sizes of 4, 8, 16, and 32 nodes correspond to MPI ranks 64, 128, 256, and 512 MPI, respectively. Containerized experiments and those with Usernetes running in the background yielded similar results. Means and standard deviations for all experiments are included in Table \ref{table:lammps}}
  \label{fig:lammps}
\end{figure}

We ran LAMMPS at a problem size of $16 \times 16 \times 8$ across job sizes of 4, 8, 16, and 32 nodes, 20 iterations each, corresponding to 64, 128, 256, and 512 MPI ranks, respectively, and across 5 setups. This produced a total of 400 individual runs. LAMMPS running on bare metal is compared to LAMMPS running in Usernetes in \ref{fig:lammps}. LAMMPS on bare metal was similar with or without Usernetes running, and running from within a container or directly on the virtual machine.  We observed LAMMPS to strong scale; LAMMPS wall times, including means and standard deviations for each scoped experiment are included in Table \ref{table:lammps}.

\begin{table*}[h]
  \centering
  \caption{LAMMPS Performance Across Environments}
  \label{table:lammps}
  \begin{tabular}{@{}p{4cm}lllll@{}}
    \toprule
    & Nodes & Ranks & Mean (s) (stddev) & \% CPU \\
    \midrule
bare-metal & 4 & 64 & 82.0 \hfill (0.0) & 99.8 \\
&  8 &  128 &  45.2 \hfill (0.616) & 99.695 \\
&  16 &  256 &  27.2 \hfill(9.15) &  99.535 \\
&  32 &  512 &  14.05 \hfill(0.224) &  99.5 \\
bare metal with usernetes & 4 &  64 &  82.1 \hfill(0.308) & 99.6 \\
&  8 &  128 & 45.75 \hfill(0.444) &  99.5 \\
&  16 & 256 & 26.0 \hfill(0.0) & 99.435  \\
&  32 & 512 & 15.85 \hfill(0.366) &  99.3 \\
container &  4 & 64 & 83.2 \hfill(5.367) &  99.78 \\
&  8 & 128 & 45.0 \hfill(0.0) & 99.7 \\
& 16 & 256 & 25.1 \hfill(0.308) &  99.595 \\
&  32 &  512 & 15.4 \hfill(5.807) & 99.495 \\
container with usernetes &  4 & 64 & 82.6 \hfill(0.503) & 99.6 \\
&  8 & 128 & 45.95 \hfill(0.224) & 99.5 \\
&  16 &  256 & 26.0 \hfill(0.0) & 99.435 \\
& 32 & 512 & 15.95 \hfill(0.224) & 99.3 \\
usernetes & 4 & 64 & 83.9 \hfill(4.025) & 99.68 \\
& 8 & 128 & 46.0 \hfill(0.0) & 99.7  \\
& 16 & 256 & 26.85 \hfill(0.366) & 99.6 \\
& 32 & 512 & 17.4 \hfill(5.798) & 99.6 \\
    \bottomrule
\end{tabular}
\end{table*}

Table \ref{table:lammps} shows nodes, ranks, times, and percentage CPU utilization for each experiment. Time (seconds) and CPU (\%) are reported with means and standard deviations. CPU utilization and mean wall times are comparable across setups, with the largest difference for the largest size of 32 nodes being approximately 3.35 seconds between bare metal (14.05 +/- 0.224) and Usernetes (17.4 +/- 5.798) seconds. 

\subsection{OSU Benchmarks}

\begin{figure}
  \centering
  \includegraphics[scale=0.58]{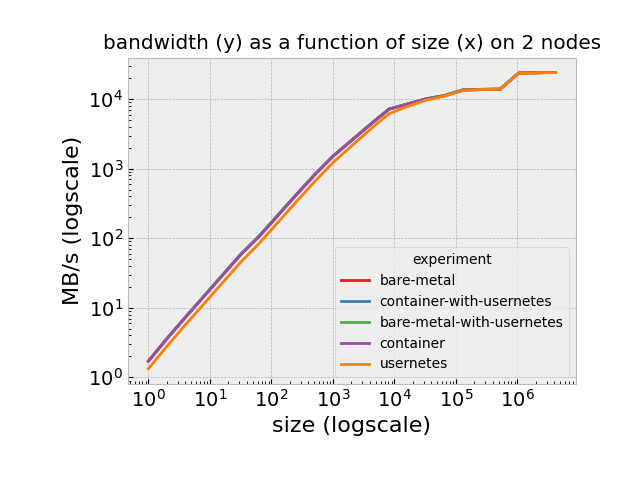}
  \caption{The OSU Bandwidth point to point benchmark. Point to point bandwidth was approximately 22.3\% lower (lower is less performant) in Usernetes as compared to bare metal for the smallest message size of 1 byte, and  0.32\% lower for the largest message size.}
  \label{fig:osu-bandwidth}
\end{figure}
\begin{figure}
  \centering
  \includegraphics[scale=0.58]{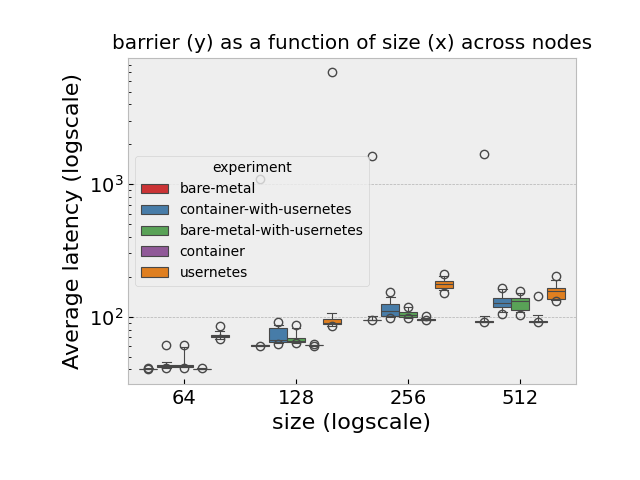}
  \caption{The OSU Barrier benchmark demonstrates Usernetes average latency is about 31.89 microseconds higher for 4 nodes (64 ranks). Having it running in the background during the execution of any bare metal (in container or not) workload introduces a detriment to average latency, whether containerized or not, and this effect is more pronounced at larger scales. }
  \label{fig:osu-barrier}
\end{figure}
\begin{figure*}
  \centering
  \includegraphics[scale=0.58]{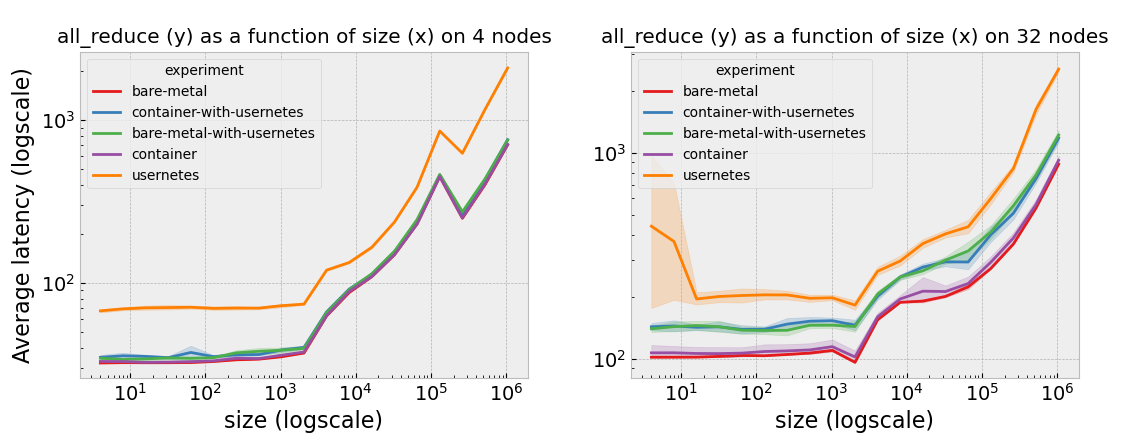}
  \caption{AllReduce latency across bytes sizes for experiments on the smallest size, 4 nodes, and the largest size 32 nodes, across message sizes. On 4 nodes Usernetes average latency was 3.6x to 13.7x greater, and for 32 nodes it was 2.89x to 4.32x greater.}
  \label{fig:osu-all-reduce}
\end{figure*}
\begin{figure}
  \centering
  \includegraphics[scale=0.58]{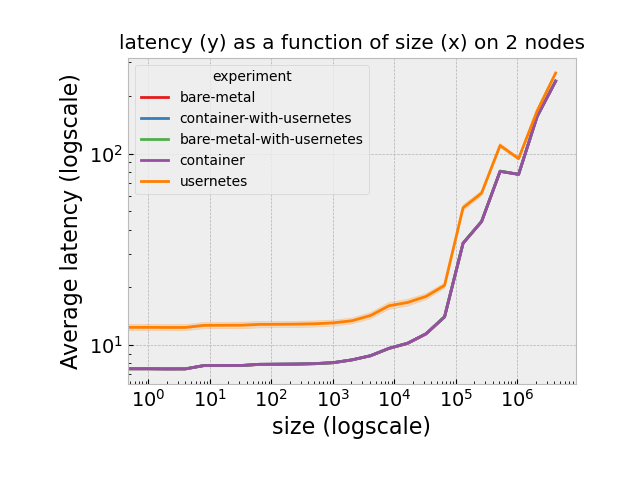}
  \caption{The OSU Latency point to point benchmark had comparable performance across ``bare-metal" setups, and 65\% greater for the smallest size in Usernetes. The ``bare-metal" (with and without container) experiments have an average latency of 7.5 microseconds, and Usernetes 12.3 microseconds.}
  \label{fig:osu-latency}
\end{figure}

We ran a subset of OSU Benchmarks across job sizes of 4, 8, 16, and 32 nodes, 20 iterations each, corresponding to 64, 128, 256, and 512 MPI ranks, respectively, and across 5 setups for each of AllReduce, Barrier, and point to point latency and bandwidth. Point to point bandwidth (osu\_bw) was approximately 22.3\% lower (meaning less data transferred) in Usernetes as compared to ``bare metal'' for the smallest message size of 1 byte (1.712 MB/s vs. 1.3 MB/s, respectively), and 0.32\% lower for the largest message size of 4,194,304 bytes (approximately 24,125 MB/s vs. 24,202 MB/s, respectively) (Figure \ref{fig:osu-bandwidth}). For OSU Latency (osu\_latency) we found that average latency for Usernetes is 65\% greater for the smallest size (Figure \ref{fig:osu-latency}) (7.46 and 12.31 microseconds, respectively). For collective call OSU Barrier (osu\_barrier) Usernetes average latency is 31.89 microseconds higher for 4 nodes (Figure \ref{fig:osu-barrier}), which means the average latency for Usernetes is 78.68\% greater. This particular benchmark had huge outliers across experiments, suggesting a common weakness that might relate to the physical location of one or more nodes. Across OSU benchmarks, bare metal performance was equivalent to performance running in a container, and the addition of Usernetes in the background added overhead that impacted the benchmarks only for collective benchmarks at larger sizes. This suggests that an allocation running Usernetes alongside bare metal HPC does incur a cost to the network, and should be considered with the setup. For the AllReduce benchmark (Figure \ref{fig:osu-all-reduce}) Usernetes average latency for 4 nodes was $3.6\times$ to $13.7\times$ higher, and for 32 nodes it was $2.89\times$ to $4.32\times$ higher (Figure \ref{fig:osu-all-reduce}). 

This observation further suggests that there is higher inter-communication cost for nodes in Usernetes, likely due to the extra overhead added by the Kubernetes headless service and other components of the software stack (Flux Framework and MPI) that might not bypass slirp4netns.

\subsection{Hybrid Demonstration of Environment Capabilities}

We ran 1000 training samples of LAMMPS using Flux with randomly chosen problem sizes ranging between 1 and 8 for each of parameters x, y, z to train three regression models served by an online machine learning server. We then ran 250 jobs of the same nature, but held out the actual wall time (Y) of the run and asked the server for a predicted value, saving the value for each request. While the specific results of the models are not a contribution of this paper, we can report that the paradigm to run HPC simulations with a workload manager alongside a Usernetes-owned service led to a successful outcome, with three different models for each of  linear, bayesian regression, and passive aggressive regression. As we are primarily interested in the hybrid environment setup, we direct the interested reader to our results repository for diagrams of the models \cite{flux-usernetes}.

\section{Discussion}

Here we presented a novel architecture to deploy user-space Kubernetes with an HPC workload manager, and ran HPC applications to define an early baseline for performance. Through our experiments, we suggest that the setup is prime to support hybrid applications that require services and HPC simulations, and that the services are well-oriented to interact with applications run directly on bare-metal. Our work demonstrates the potential of this setup for hybrid workflows that require both the performance of HPC and modular service-oriented architecture of Kubernetes, and also suggest that further work is needed to better understand the remaining gap in performance between the setups, and to develop orchestration tools and meta-schedulers that afford easy interaction between the spaces. While this early work was performed in a cloud environment for reproducibility and sharing with the larger community, we imagine this hybrid setup to ultimately be most useful for on premises systems, offering the best of both worlds from the cloud and HPC communities. We provide the community with a reproducible deployment to be empowered to further work on this setup with us \cite{flux-usernetes}.

Despite the network performance not being equal between Usernetes and ``bare metal'' setups, we were surprised that the LAMMPS application had similar performance, and that differences in means were only a single order of magnitude, as opposed to $2\times$ slower, which is what we saw in a similar virtual machine setup that used ethernet as the underlying network \cite{fosdem-bare-metal-bros}. We believe this difference comes down to the likelihood that this particular build of LAMMPS is CPU bound at this smaller size, and using the elastic fabric adapter (EFA) vs. standard ethernet. The EFA verbs devices that are mounted into the container ignore the kernel entirely, and thus are not subject to the additional filtering of the TAP device. Messages over IBVerbs reach the NIC directly, implying that messages over TCP might be subject to the extra processing by the TAP device, but RDMA is not. The loss in performance may be due to other factors that warrant further study. A first area of investigation is the Kubernetes headless Service that provides each of the Pods running Flux with a hostname. This hostname needs to be discovered by traversing a local lookup table, and this might be adding overhead. It is also possible that other components of the software stack included with Flux or associated MPI libraries might not be bypassing slirp4netns. We anticipate further work in this area to continue to improve upon Usernetes' ability to offer low latency.

While the paradigm of an HPC workload manager running alongside infrastructure like Kuberneters is new, we can anticipate several exciting, possible future developments that will further support this setup. First, Kubernetes has support for custom schedulers. A current issue with the setup described here is that each of Usernetes and Flux Framework has their own scheduler, and we handled this issue by scheduling Usernetes within a Flux instance. The was worked around largely by requesting jobs that leave room for the other scheduler, however a more ideal approach would use a single scheduler that can account for scheduling to both resources, and this would be possible with a custom scheduler framework. Our group has done work in this space \cite{Misale2021-tv,Milroy2022-pv} and will extend the work further. A second area for development that is specific to Flux Framework is to make it easy to submit workloads with logic that can orchestrate between the two environments. For example, if we take our hybrid machine learning prototype and extend it to be across node sizes and types (e.g., with or without GPU), we would want to use a Flux batch job with nested brokers that each control a scheduler for each group of resources. This is one of the core design features of Flux that was not used here, and we believe it would contribute to the flexibility and efficiency of the design.

We learned a great deal about good practices for deploying this kind of architecture in the AWS EC2 cloud, and specifically, attributes related to running user space Kubernetes alongside a workload manager on bare metal. First, while Pod resource requests and limits can be set to ensure a one to one mapping of Pods to nodes, and while the number of processes and cores might appear to be consistent between the virtual machine and a deployed Pod by using a tool like hwloc \cite{hwloc}, this does not mean that the pod is able to fully utilize all resources. While we thought that a Pod in ``Burstable" mode \cite{qos} could go over the resource limit, in practice this didn't turn out to be the case. We discovered that setting a resource limit of CPU did not limit the CPU count, but did set a maximum for the percentage CPU cycles allowed, which can reduce an application's CPU utilization from 99.9\% down to 40-60\%, which can have huge implications for performance. We encourage developers to only specify the EFA device in the resource requests and limits, and to use affinity to map one pod per node without impacting performance.

Next, it is important to be aware that systemd services can silently set limits on the workload manager's ability to utilize memory or other system resources. If issues arise, we suggest investigating this setup. And finally, while outside of the scope of this work, we believe that other approaches that bypass the OS for an application to communicate with a network device should be explored, including (but not limited to) Infiniband and other network drivers.

We also point to this project as a converged computing success in terms of collaboration. Developers from the HPC community, the authors of this paper, successfully collaborated with cloud native projects to push forward an important update to the Usernetes project that will likely further empower both sides. In the same vein, this work is an example of the power of small numbers of people. The Usernetes project is maintained, at most, by a handful of developers, and the authors of this paper were also small in numbers. However, through communication and advocacy, a novel paradigm was imagined and developed, and we believe it can have an impact on the larger HPC technology space. We encourage the larger community to also pursue ideas they believe in, and not be discouraged by their own sense of importance or numbers.

\begin{IEEEbiographynophoto}{Vanessa Sochat} is a Computer Scientist in Livermore Computing. Her work focuses on container technologies and high performance computing convergence in cloud. She maintains the Flux Operator and Fluence custom scheduler as part of the Flux Framework, and leads an ISCP for User-space Kubernetes. She collaborates closely with industry and several open source communities. She received her Ph.D and M.S. in Biomedical Informatics from Stanford University, and her B.A. from Duke University.
\end{IEEEbiographynophoto}

\begin{IEEEbiographynophoto}{David Fox} is a Computer Scientist in the Livermore Computing division at Lawrence Livermore National Laboratory, where his work involves, HPC, Cloud and Storage systems. He is a leader for the HPC Cluster Engineer Academy. He holds a B.S. in Electrical Engineering from Brigham Young University.
\end{IEEEbiographynophoto}

\begin{IEEEbiographynophoto}{Daniel Milroy} is a Computer Scientist at the Center for Applied Scientific Computing. His research focuses on graph-based scheduling and dynamic resource management for high-performance computing (HPC) and cloud-converged environments, and he collaborates closely with industry and academia. Dan is a developer of the Fluxion directed graph-based scheduler in the Flux Framework and leads a thrust of the Fractale LDRD. Dan holds a Ph.D. and M.S. in Computer Science from the University of Colorado Boulder, and an A.B. in Physics from the University of Chicago.
\end{IEEEbiographynophoto}

\section*{Acknowledgments}
Author VS: Thank you to the larger HPC community for interesting discussion on these setups, but especially to Mark Grondona in the Flux Framework team that identified the last missing detail of the setup - a memory limit set by systemd that I never imagined could be a thing in my wildest dreams. This work was performed under the auspices of the U.S. Department of Energy by Lawrence Livermore National Laboratory under Contract DE-AC52-07NA27344 (LLNL-JRNL-865427-DRAFT) and was supported by the LLNL-LDRD Program under Project No. 22-ERD-041.
\bibliographystyle{IEEEtran}
\bibliography{references}

\begin{thebibliography}{10}
\bibitem{flannel}
``\BIBforeignlanguage{en}{flannel: flannel is a network fabric for containers, designed for kubernetes}.''
\end{thebibliography}

\newpage

\vfill

\end{document}